\documentclass{llncs}

\usepackage{amsmath}
\usepackage{graphicx}

\begin{document}

\title{Combinatorial Miller-Hagberg Algorithm for Randomization of Dense Networks}

\author{Hiroki Sayama \inst{1,2}}
\institute{Center for Collective Dynamics of Complex Systems and Department of Systems Science and Industrial Engineering, Binghamton University, Binghamton, New York 13902, USA
\and
Center for Complex Network Research, Northeastern University, Boston, Massachusetts 02115, USA\\
\email{sayama@binghamton.edu}
}

\maketitle

\setcounter{footnote}{0}

\begin{abstract}
We propose a slightly revised Miller-Hagberg (MH) algorithm that
efficiently generates a random network from a given expected degree
sequence. The revision was to replace the approximated edge
probability between a pair of nodes with a combinatorically calculated
edge probability that better captures the likelihood of edge presence
especially where edges are dense. The computational complexity of this
combinatorial MH algorithm is still in the same order as the original
one. We evaluated the proposed algorithm through several numerical
experiments. The results demonstrated that the proposed algorithm was
particularly good at accurately representing high-degree nodes in
dense, heterogeneous networks. This algorithm may be a useful
alternative of other more established network randomization methods,
given that the data are increasingly becoming larger and denser in
today's network science research.
\end{abstract}

\section{Introduction}

In network science, there are occasions in which one needs to generate
random network samples from a given node degree sequence. A typical
context for doing this is to conduct a statistical test of whether
empirically observed network properties can be explained by a certain
degree distribution or not. Several algorithms have already been
developed for this purpose, such as the classic Havel-Hakimi algorithm
\cite{hakimi1962realizability}, the double edge swap method, the
configuration model \cite{bender1978asymptotic,newman2003structure}, and the
Bayati-Kim-Saberi algorithm \cite{bayati2010sequential}. However, they
come with respective limitations. The Havel-Hakimi algorithm
constructs a network using a heuristic, assortativity-inducing
procedure, whose outcomes would not be appropriate to be used as fully
randomized controls. The double edge swap method is simple but its
randomization process is slow and gradual, with no well-defined
termination condition. The configuration model is a systematic,
well-defined randomization method, but its outcomes often contain
parallel edges and self-loops. The Bayati-Kim-Saberi algorithm can be
computationally costly and does not guarantee that it can produce a
randomized graph as an output.

The Miller-Hagberg algorithm \cite{miller2011efficient} (called the MH
algorithm hereafter) addresses those limitations of the other
algorithms mentioned above by relaxing the requirement so that it
generates a random network from a given {\em expected} degree
sequence. This relaxation allows for calculation of edge probability
{\em independently} for each pair of nodes. By sorting the nodes
according to their expected degrees and implementing an efficient
node-skipping mechanism (see \cite{miller2011efficient} for details),
the MH algorithm achieves linear computational complexity $O(N+M)$,
where $N$ and $M$ are the numbers of nodes and edges,
respectively. This property is highly desirable for large-scale
network analysis.

While the MH algorithm can be used with any edge probability
functions, its original version uses Chung and Lu's random graph model
\cite{chung2002connected} that assumes that an edge probability
between two nodes with degrees $w_i$ and $w_j$ can be approximated as
$\min(1, w_iw_j / \sum_k w_k)$. It is known that this assumption is invalid if
the network is dense (i.e., if $w_i$ is not negligible compared to
$N$). This issue is typically manifested on high-degree nodes whose
degrees generated by this algorithm often deviate greatly from their
expected degrees specified in the given degree sequence
\cite{miller2011efficient}\footnote{There have been a couple of
  modifications of edge probability calculation proposed to address
  this issue \cite{britton2006generating,van2013critical}, mostly
  using statistical physics approaches.}. This limitation has not been
so critical an issue so far because most real-world networks show
significant degree heterogeneity and thus they are fundamentally
sparse \cite{del2011all}.

With the recent expansion of modeling methodologies and application
domains of network science, however, there are now several situations
in which one needs to analyze {\em dense} networks, such as the ego
networks in social media data \cite{leskovec2012learning}, the
time/layer aggregations of temporal and multilayer networks
\cite{holme2012temporal,kivela2014multilayer}, and the functional
connectivity networks of the brain imaging data \cite{zamani2017}, to
name a few. These networks typically have much higher edge density
than other more classical networks, while they still maintain
substantial degree heterogeneity. Accurately representing their
high-degree nodes in randomized counterparts is thus an important
methodological challenge.

In this paper, we aim to address the above challenge by implementing a
small yet unique revision in the original MH algorithm, by replacing
the Chung-Lu edge probability with a combinatorically calculated edge
probability that better captures the likelihood of edge presence
especially where edges are dense. In the rest of the paper, we
describe technical details of the algorithm revision and then present
some results of evaluation of the proposed algorithm through numerical
experiments.

\section{Revising the MH Algorithm with Combinatorial Edge Probability}

\begin{figure}[t]
\centering
\includegraphics[width=\columnwidth]{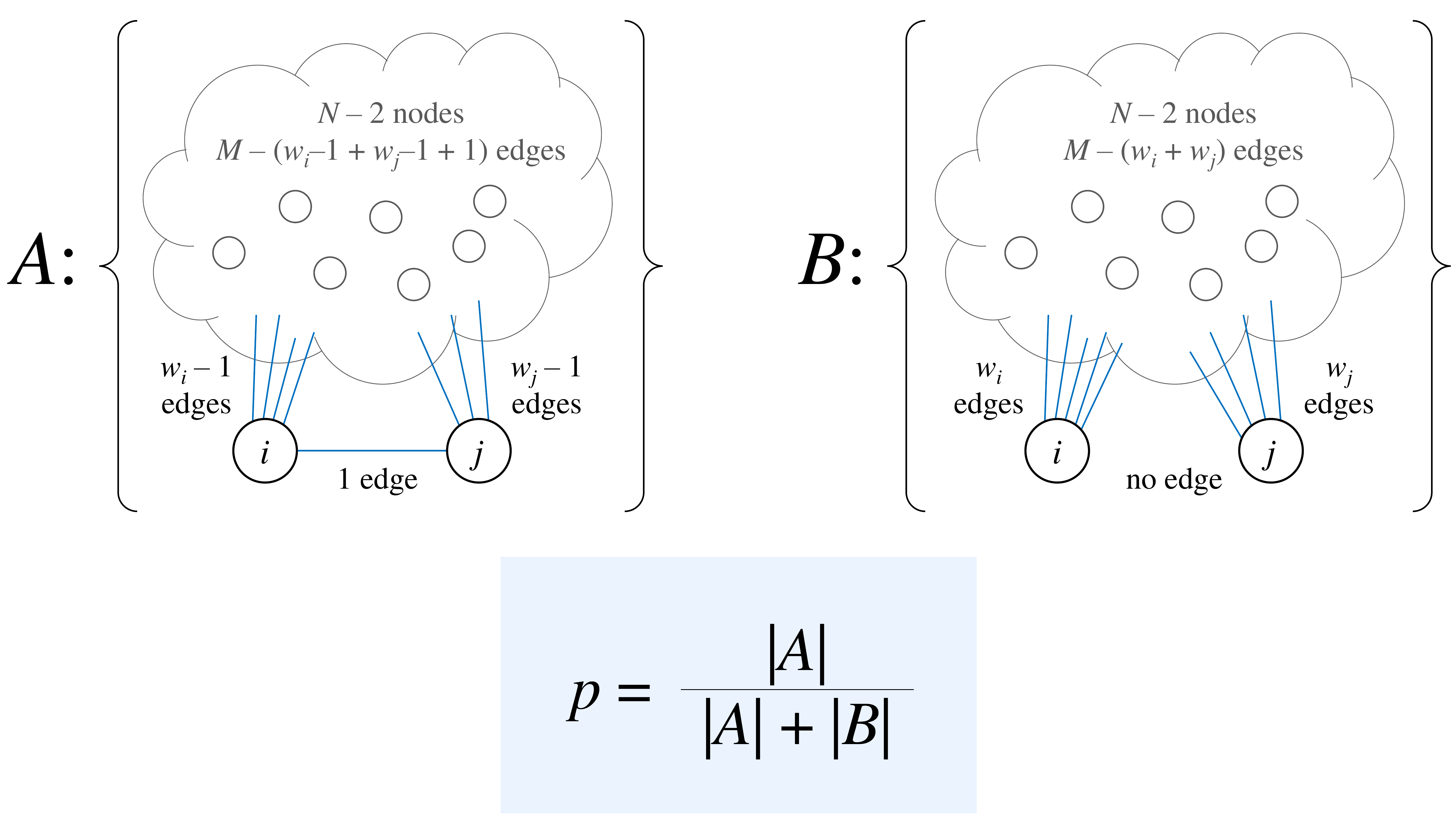}
\caption{Schematic illustration of the proposed combinatorial edge
  probability calculation (Eq.~\ref{newp}). $|A|=C_c(N, M, w_i, w_j)$:
  Number of network configurations in which the two focal nodes, $i$
  and $j$, are connected directly. $|B|=C_d(N, M, w_i, w_j)$: Number
  of network configurations in which the two nodes are not connected
  directly.}
\label{idea}
\end{figure}

We revise the MH algorithm by replacing the approximated edge
probability with a combinatorically calculated edge probability. This
calculation is done by counting the number of all network
configurations in each of the two scenarios: the presence or the
absence of an edge between two focal nodes. Let $w_i$ and $w_j$ be the
degrees of two nodes, $i$ and $j$, for which the edge probability
between them is to be calculated. Also let $N$ and $M$ be the numbers
of nodes and edges in the network, respectively. Assuming that each
network configuration occurs randomly with equal probability, the edge
probability between the two nodes can be written as
\begin{align}
p(N, M, w_i, w_j) &= \frac{C_c(N, M, w_i, w_j)}{C_c(N, M, w_i, w_j) + C_d(N, M, w_i, w_j)} , \label{newp}
\end{align}
where $C_c(N, M, w_i, w_j)$ is the number of network configurations in
which the two nodes $i$ and $j$ are connected directly, and $C_d(N, M,
w_i, w_j)$ the number of network configurations in which those nodes
are {\em not} connected directly (Fig.~\ref{idea}). Eq.~(\ref{newp})
can be rewritten as
\begin{align}
p(N, M, w_i, w_j) &= \left( 1 + \frac{C_d(N, M, w_i, w_j)}{C_c(N, M, w_i, w_j)} \right)^{-1}\label{newp2}
\end{align}
if $C_c(N, M, w_i, w_j) \ne 0$.

Both $C_c$ and $C_d$ can be calculated as the product of the following
three combinatorial quantities (Fig.~\ref{idea}):
\begin{itemize}
\item Number of possibilities of placing the edges that emanate from
  node $i$ to the rest of the network
\begin{itemize}
\item For $C_c$: $\displaystyle\binom{N-2}{w_i-1}$
\quad \quad\quad \quad\quad \quad\,
For $C_d$: $\displaystyle\binom{N-2}{w_i}$
\end{itemize}
\item Number of possibilities of placing the edges that emanate from
  node $j$ to the rest of the network
\begin{itemize}
\item For $C_c$: $\displaystyle\binom{N-2}{w_j-1}$
\quad \quad\quad \quad\quad \quad\,
For $C_d$: $\displaystyle\binom{N-2}{w_j}$
\end{itemize}
\item Number of possibilities of placing the edges not adjacent to the
  two nodes among the rest of nodes in the network
\begin{itemize}
\item For $C_c$: $\displaystyle\binom{\binom{N-2}{2}}{M-w_i-w_j+1}$
\quad \quad
For $C_d$: $\displaystyle\binom{\binom{N-2}{2}}{M-w_i-w_j}$
\end{itemize}
\end{itemize}
By multiplying these three quantities, we obtain
\begin{align}
C_c(N, M, w_i, w_j) &= \binom{N-2}{w_i-1}\binom{N-2}{w_j-1} \binom{\binom{N-2}{2}}{M-w_i-w_j+1}, \text{ and}\\
C_d(N, M, w_i, w_j) &= \binom{N-2}{w_i} \binom{N-2}{w_j}\binom{\binom{N-2}{2}}{M-w_i-w_j}.
\end{align}
By applying these combinatorial calculations into Eq.~(\ref{newp2}) and
simplifying it, we obtain 
\begin{align}
p(N, M, w_i, w_j) &= \left( 1 + \frac{N-w_i-1}{w_i} \frac{N-w_j-1}{w_j}
\frac{M-w_i-w_j+1}{\binom{N-2}{2} - M + w_i+w_j} \right)^{-1}\\
&= \left( 1 + \frac{2M^*(N-w_i-1)(N-w_j-1)}{w_iw_j (N^2-5N+8-2M^*)} \right)^{-1} ,
\end{align}
where $M^* = M-w_i-w_j+1$. In actual computation of $p$, we use the
following more straightforward formula that does not involve inversion:
\begin{align}
p(N, M, w_i, w_j) &= \frac{X} {X + Y} \label{newp-detail1}\\
X &= w_iw_j (N^2-5N+8-2M^*) \\
Y &= 2M^*(N-w_i-1)(N-w_j-1) \label{newp-detail2}
\end{align}
This correctly gives $p=0$ if $w_i$ or $w_j = 0$, which is convenient
for practical purposes.

The formula obtained above is surprisingly simple, involving only a
finite, constant number of basic arithmetic operations. Therefore, the
revised MH algorithm with this combinatorial edge probability (called
the {\em combinatorial MH algorithm} hereafter) still maintains the
original computational complexity $O(N+M)$. Also note that
Eqs.~(\ref{newp-detail1})--(\ref{newp-detail2}) recovers the
original Chung-Lu formula $w_iw_j / (2M) = w_iw_j / \sum_k w_k$, if $N
\to \infty$ and $w_i, \; w_j \ll M \ll N^2$.

Eqs.~(\ref{newp-detail1})--(\ref{newp-detail2}) capture the edge
probability more accurately where edge density is high. Considering
some extreme cases helps illustrate this benefit. For example, in a
complete graph made of $N$ nodes, each node has $N-1$ as its degree,
and the total number of edges is $N(N-1)/2$. Letting $w_i=w_j=N-1$ and
$M=N(N-1)/2$ (i.e., $M^*=N(N-1)/2-(N-2)-(N-2)+1$) in
Eqs.~(\ref{newp-detail1})--(\ref{newp-detail2}) produces $p=1$,
correctly indicating that any pair of nodes must be connected
directly. However, the Chung-Lu model gives $p=(n-1)/n < 1$ in the same situation. A more
extreme case is a star graph made of $N$ nodes and $N-1$ edges. The
edge probability between the central node (with $w_i=N-1$) and a
peripheral node (with $w_j=1$) is correctly calculated to be $p=1$ by
Eqs.~(\ref{newp-detail1})--(\ref{newp-detail2}), while the Chung-Lu
model gives $p=1/2$, which is far off the actual probability
1. Finally, another example that shows the opposite way of deviation
is a disconnected graph made of two 6-node star graphs ($N=12$,
$M=10$). In this graph, the edge probability between the two central
nodes ($w_i=w_j=5$) is calculated to be $p=125/129$ by
Eqs.~(\ref{newp-detail1})--(\ref{newp-detail2}), which correctly
captures the small possibility that those two central nodes do not
have a direct connection to each other. In the meantime, the Chung-Lu
model gives $p=\min(1, 5/4) = 1$, which forces the two central nodes to {\em
  always} be connected in randomized networks. These examples
demonstrate the accuracy of the combinatorial edge probability
proposed in this study.

We note that Eqs.~(\ref{newp-detail1})--(\ref{newp-detail2}) may not
provide accurate edge probabilities for low-degree nodes. For example,
they give a non-zero (positive) edge probability between two
peripheral nodes in a star graph, since their mandatory connections to
the central node are ignored when the edge probability between them is
calculated. In general, the proposed algorithm tends to produce
slightly higher-than-expected degrees for peripheral nodes in
heterogeneous networks (which will be seen in numerical results
later). Also, Eqs.~(\ref{newp-detail1})--(\ref{newp-detail2}) may
malfunction if a graphically impossible input is given, because the
formula was derived using combinatorial enumerations under the
assumption that the given parameters ($N$, $M$, $w_i$, $w_j$) are
graphically possible. For example, $(N, M, w_i, w_j) = (5, 10, 1, 1)$
(which is graphically impossible) gives a meaningless value $p =
-5/76$. However, such a problem will not arise as long as the formula
is used for randomizing the topology of an existing network. In what
follows, we exclusively consider cases in which the expected degree
sequence is always obtained from the degree sequence of another
existing network.

\section{Evaluations}

We first tested the proposed combinatorial MH algorithm by applying it
to several illustrative dense networks. The following three networks
were used:
\begin{itemize}
\item Zachary's Karate Club network \cite{zachary1977information} (34
  nodes; 78 edges; density: 0.139)
\item Ego network of an arbitrarily chosen user (user `3000' for this
  example) in Leskovec-McAuley Facebook dataset
  \cite{leskovec2012learning} (92 nodes; 2,044 edges; density: 0.488)
\item Dense heterogeneous network constructed using the
  Barab\'{a}si-Albert model \cite{barabasi1999emergence} (300 nodes;
  20,000 edges; density: 0.446)
\end{itemize}

\begin{figure}
\centering
\includegraphics[height=0.28\textheight,width=0.8\columnwidth]{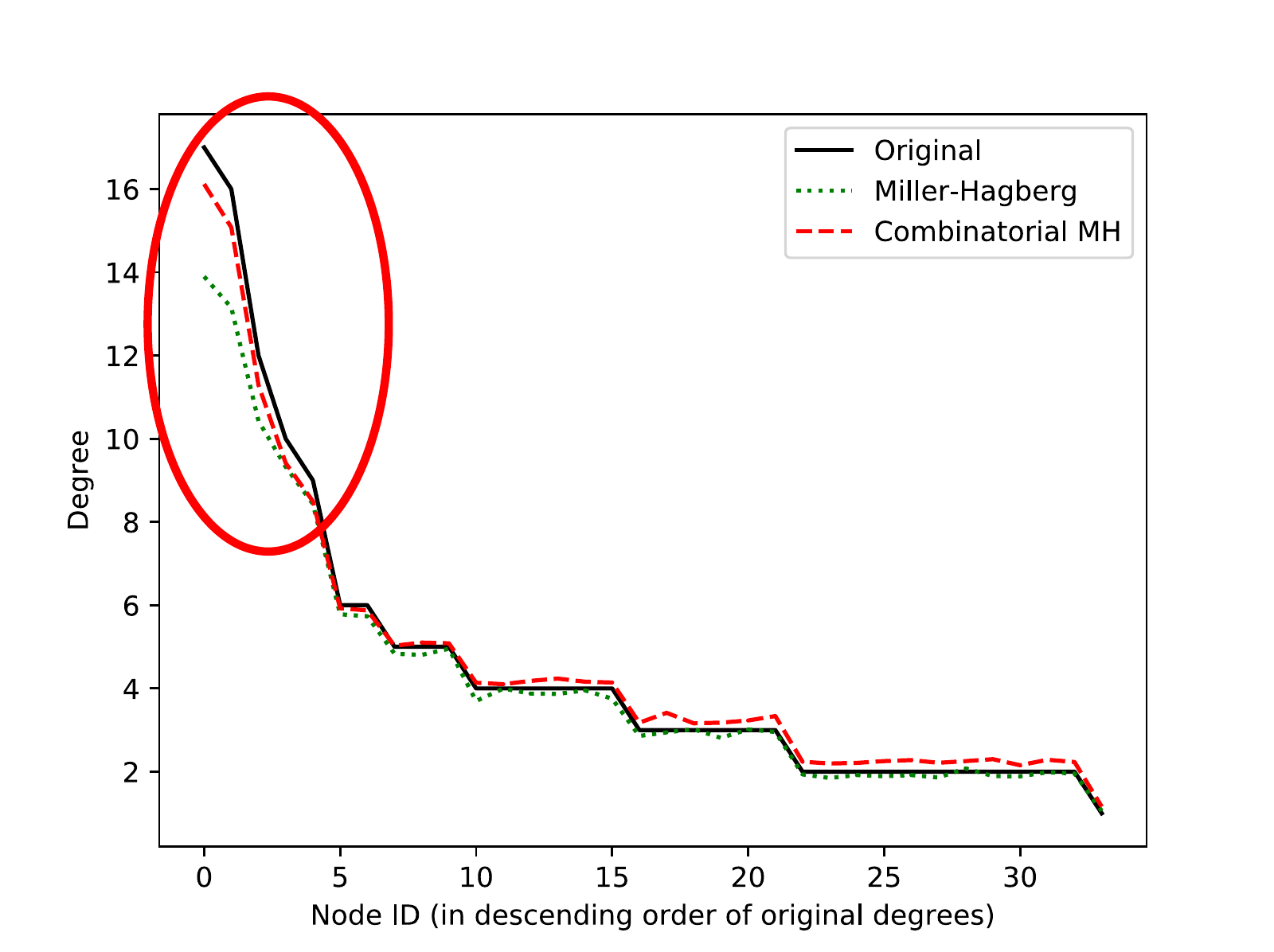}\\
\includegraphics[height=0.28\textheight,width=0.8\columnwidth]{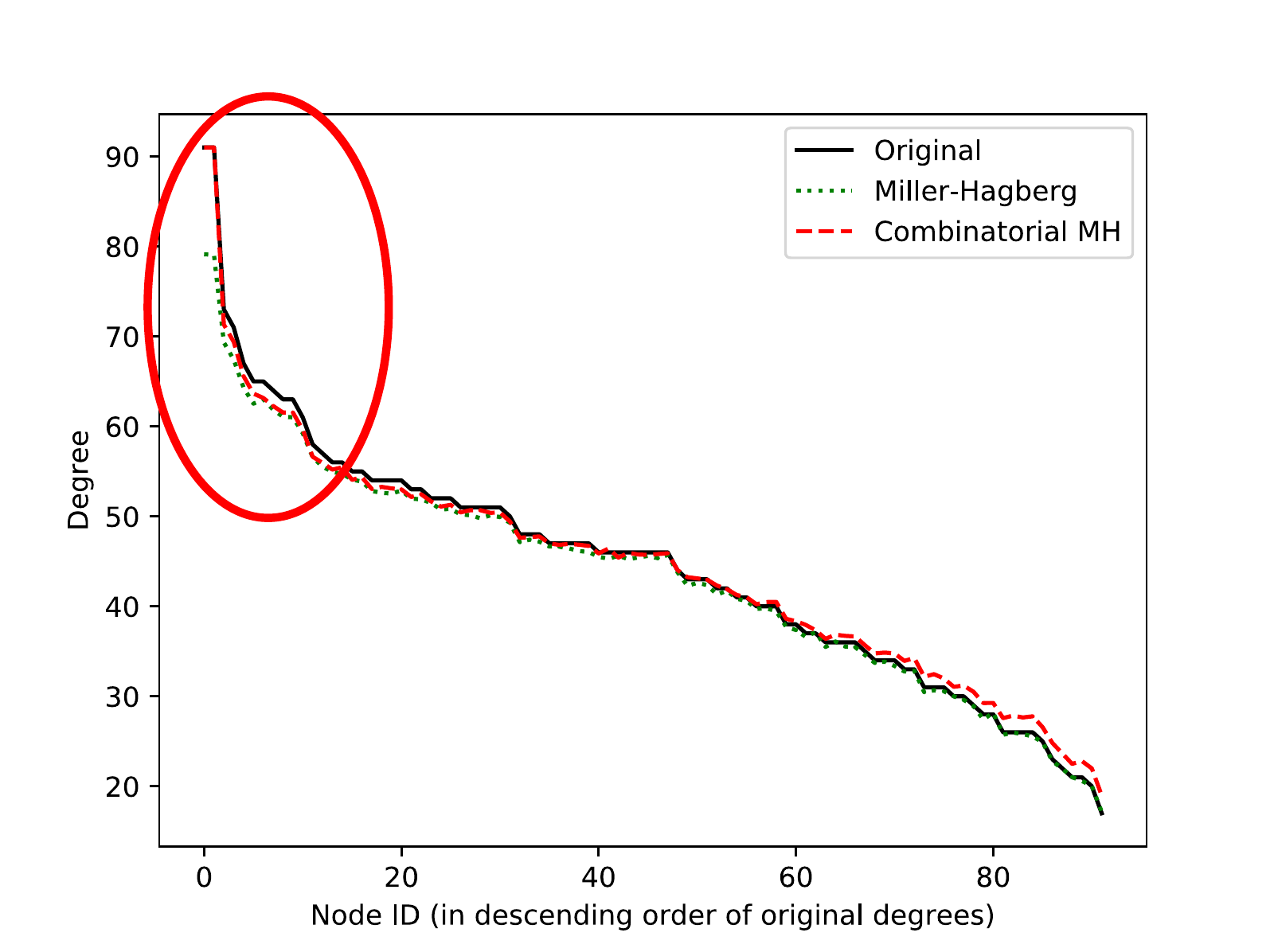}\\
\includegraphics[height=0.28\textheight,width=0.8\columnwidth]{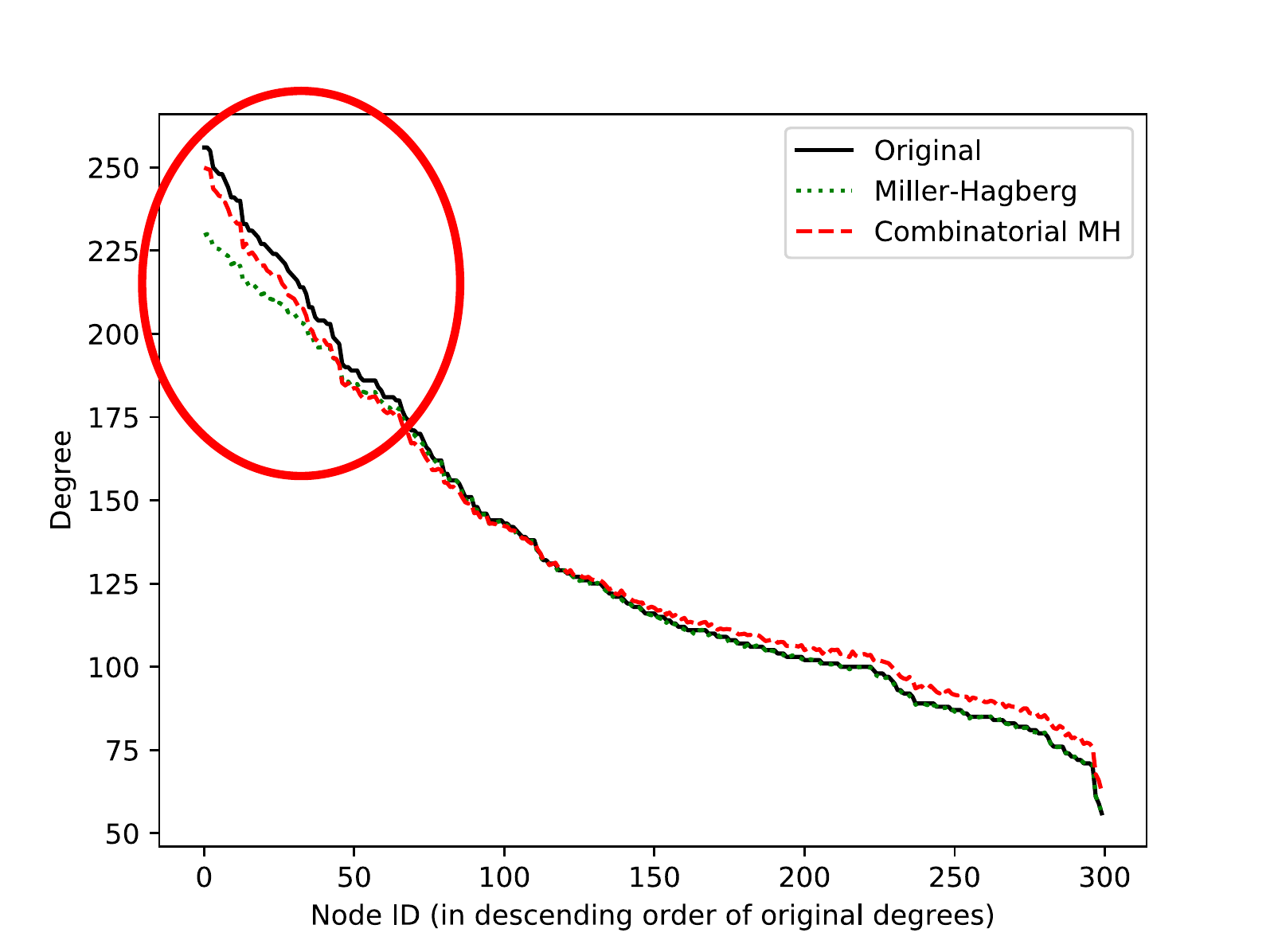}
\caption{Comparison of degree sequences among the original network
  (black, solid lines) and two randomized ones (green, dotted lines:
  original MH algorithm; red, dashed lines: combinatorial MH
  algorithm). Top: Zachary's Karate Club network
  \cite{zachary1977information}. Middle: Ego network in
  Leskovec-McAuley Facebook dataset
  \cite{leskovec2012learning}. Bottom: Dense heterogeneous network
  constructed using the Barab\'{a}si-Albert model
  \cite{barabasi1999emergence}. For each randomization algorithm, the
  average result of 500 independent randomization trials is
  shown. Nodes are sorted in descending order of their degrees in the
  original network. A clear difference between the original and
  combinatorial MH algorithms is seen on high-degree nodes
  (highlighted with red circles).}
\label{results1}
\end{figure}

Figure \ref{results1} shows the results in which the degree sequences
among the given original network and two randomized ones (by the
original and combinatorial MH algorithms) were compared. For each
randomization algorithm, the average result of 500 independent
randomization trials is shown. It is clearly seen in these plots that
the combinatorial MH algorithm (red, dashed lines) was able to
represent high-degree nodes more accurately than the original MH
algorithm (green, dotted lines).

\begin{figure}[t]
\includegraphics[width=0.5\columnwidth,height=0.4\columnwidth]{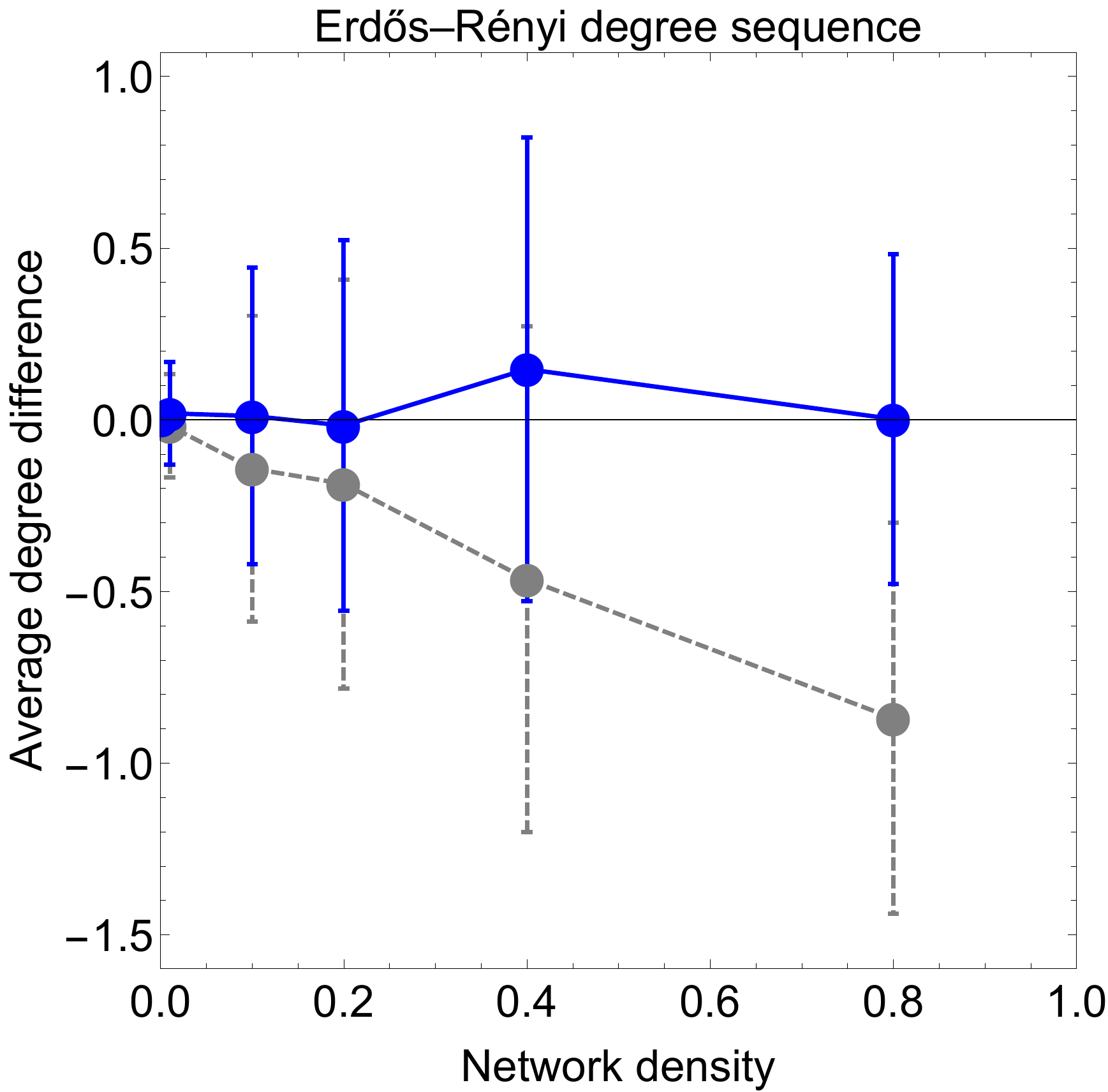}
\includegraphics[width=0.5\columnwidth,height=0.4\columnwidth]{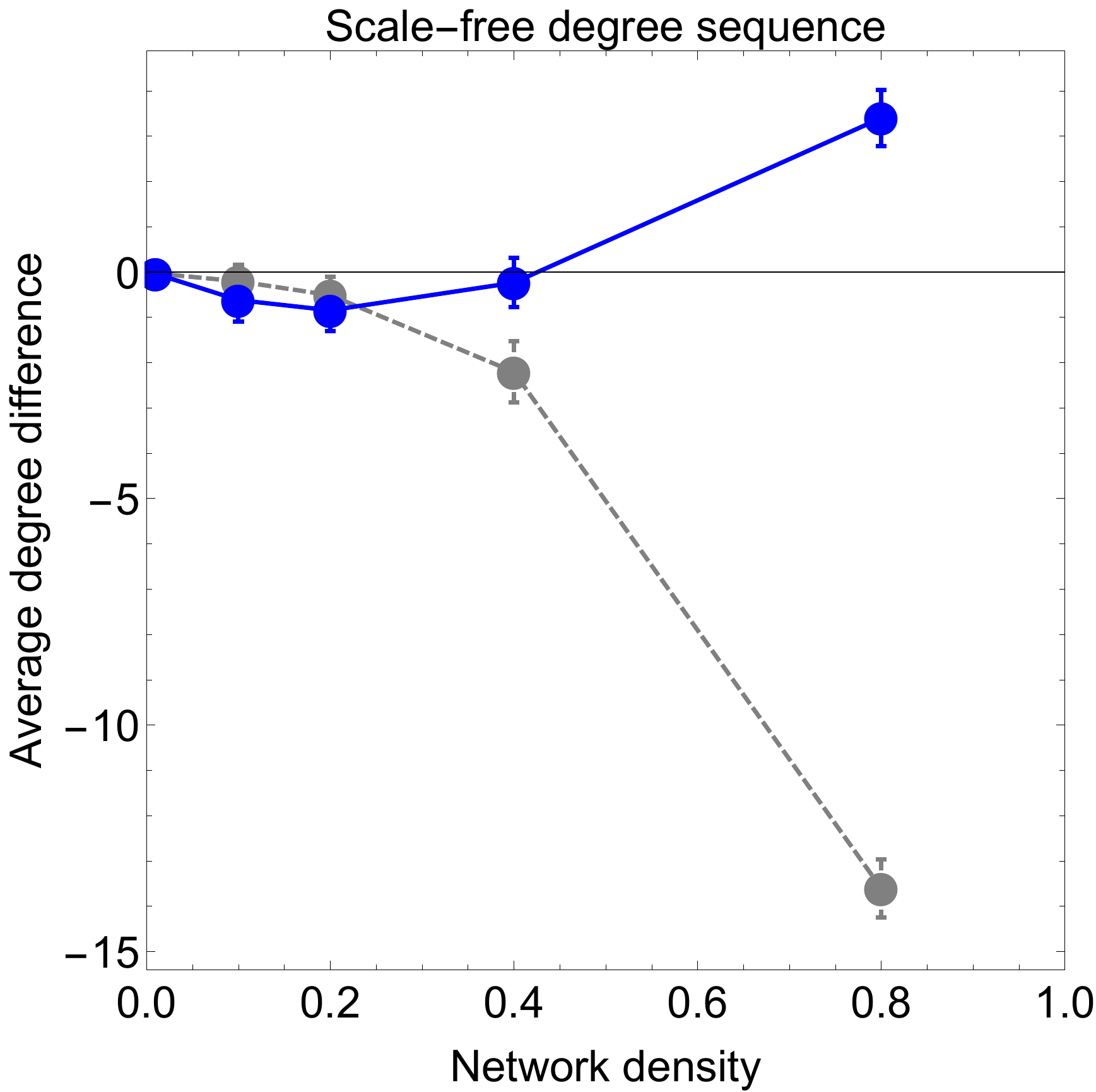}
\caption{Performance comparison between the original (gray, dashed
  lines) and combinatorial (blue, solid lines) MH algorithms. Each
  algorithm was applied to a randomly generated expected degree
  sequence (left: sequence generated from Erd\H{o}s-R\'enyi networks,
  right: sequence generated from Barab\'{a}si-Albert networks) over
  varying network densities. $N=1,000$ for all cases. Performance was
  measured by the difference in average node degrees between given and
  randomized networks. Each data point was an average of 100
  independent simulations. Error bars show standard
  deviations. Similar trends were observed for $N = 100$, $500$ and
  $2,000$.}
\label{results2}
\end{figure}

We also evaluated the effect of edge density on the performance of
randomization algorithms. Figure \ref{results2} shows the results of a
numerical experiment in which the edge density was systematically
varied on Erd\H{o}s-R\'enyi and Barab\'{a}si-Albert networks. The
performance of the algorithms was measured by the difference in
average node degrees between given and randomized networks. The
combinatorial MH algorithm successfully reproduced average node
degrees that were closer to the given ones, especially for high edge
density cases.

\section{Conclusions}

In this paper, we presented the combinatorial MH algorithm in which
the edge probability between a pair of nodes was combinatorically
calculated. The derived edge probability formula involved only a
constant number of basic arithmetic operations, keeping the linear
computational complexity of the original MH algorithm. Numerical
experiments demonstrated that the proposed algorithm was particularly
good at accurately representing high-degree nodes in dense,
heterogeneous networks. This algorithm may be a useful alternative of
other more established network randomization methods, given that the
data are increasingly becoming larger and denser in today's network
science research.

What is particularly unique about the proposed algorithm is that it
captures, in some sense, certain non-local topological dependencies in
calculating edge probability (this helps accuracy), even though the
probability itself is still calculated independently for each node
pair (this helps computational efficiency). In the meantime, such
independent calculation of edge probability may also be a limitation
of the algorithm because, as noted earlier, it may produce inaccurate
results where edges are sparse. This limitation should be taken into
account when one decides which network randomization algorithm should
be used for a specific network dataset. Proper handling of such
interdependency of edge probabilities will require more careful
mathematical analysis and algorithm design, which is among our future
work.

\bibliographystyle{splncs}
\bibliography{sayama}

\end{document}